\documentclass[aps,amssymb,twocolumn]{revtex4}
\usepackage{amssymb}
\usepackage{amsmath}
\usepackage{graphicx}
\usepackage{color}
\usepackage{textcomp}
\usepackage{ulem}
\usepackage{lipsum}
\begin{document}

\title{  Coupling strength induced BCS-BEC crossover on phase boundary of pion superfluid }
\author{Zhiyang Liu}
\author{Shijun Mao }
 \email{maoshijun@mail.xjtu.edu.cn}
\affiliation{School of Physics, Xi'an Jiaotong University, Xi'an, Shaanxi 710049, China}

\begin{abstract}
Coupling strength effect on the quark matter with finite isospin chemical potential is studied in a Pauli-Villars regularized NJL model. A BCS-BEC crossover occurs along the phase boundary of pion superfluid phase transition, as increasing coupling strength $G$. For strong coupling cases, the critical isospin chemical potential for pion superfluid phase transition $\mu_I^c$ is exactly the same as pion mass in vacuum $M_\pi$. Around the critical point $\mu_I^c$, the pion superfluid quark matter is in BEC state, associated with a fast increase of pion condensate. For weak coupling cases, we obtain $\mu_I^c<M_\pi$, and a mass jump of the Goldstone boson at the critical point $\mu_I^c$. The pion superfluid quark matter is in BCS state even around $\mu_I^c$, accompanied by a slow increase of pion condensate. Note that $\mu_I^c$ is a non-monotonic function of coupling strength $G$. On the other hand, coupling strength effect changes the bulk properties of quark matter. In strong (weak) coupling cases, the EoS of quark matter at finite isospin chemical potential is stiff (soft). Therefore, the compact stars composed of strong (weak) coupling pion superfluid quark matter has a heavier (lighter) mass and larger (smaller) radius.
\end{abstract}

\date{\today}
\maketitle

\section{Introduction}

There are two kinds of condensed states in a typical fermion gas: the Bardeen-Cooper-Shrieffer condensation (BCS) of fermion pairs where the pair size is large and the pairs overlap each other, and the Bose-Einstein condensation (BEC) of molecules where the pair size is small and the pairs are distinguishable. Although the BCS and BEC limits are physically quite different, the change from BCS to BEC was found to be
smooth. On theoretical sides, the BCS wave function can be generalized to arbitrary attraction which leads to a smooth crossover from BCS to BEC~\cite{bcsbec1,bcsbec2,bcsbec21,bcsbec22}. On experimental sides, the BCS-BEC crossover in cold atom systems has been realized through the technic of Feshbach resonance~\cite{wx6,wx7,wx8,wx9,bcsbec3,bcsbec4,bcsbec5,bcsbec6,bcsbec7}, which can be used to tune the interaction between fermions.

The BCS-BEC crossover has also been studied in quark matter. The increasing isospin (baryon) density leads to a phase transition from normal quark matter to a pion superfluid (color superconductor) due to the spontaneous breaking of symmetry~\cite{cscpi1,cscpi2,cscpi3}. Inside the pion superfluid (color superconductor) phase, the BCS-BEC crossover happens with the change of isospin (baryon) density~\cite{reviewbcs1,reviewbcs2}, which is described by the quark chemical potential, the size of Cooper pair, the pion (diquark) condensate and pair scattering length. Previous studies focus on the density induced BCS-BEC crossover in quark matter. By analogy with the usual fermion gas, there will be the coupling strength triggered BCS-BEC crossover in quark systems, which is rarely studied. In our current work, we will investigate the coupling strength induced BCS-BEC crossover in pion superfluid quark matter.

The study of phase transitions at moderate temperature and isospin density relies on lattice QCD calculations and effective models with QCD symmetries. One of the widely used effective models is the Nambu-Jona-Lasinio (NJL) model~\cite{njl1,njl2,njl3,njl4,njl5}, which was originally inspired by the BCS theory. Its version at quark level gives a simple and direct description of the spontaneous breaking of chiral and isospin symmetries, and describes the static properties of light mesons remarkably well.

The change of coupling strength in the quark system will not only lead to different phase structure, but also modify the bulk properties of the system, such as the equation of state (EoS). The quantum chromodynamics (QCD) phases at high density might be realized in the core
of compact stars~\cite{wx21,wx22,comp1,comp2,comp3}, and the structure of compact stars is closely related to the state of quark matter inside. The mass-radius relation of compact stars composed of pion superfluid quark matter will be calculated by considering different EoS with different coupling strength.

This paper is organized as follows. In Sec.\ref{sec:f}, we briefly review the NJL model at finite isospin chemical potential. The phase structure of quark matter, mass spectra of mesons and the mass-radius relation of the corresponding compact stars are discussed in Sec.\ref{sec:r}, and we focus on the coupling strength effect. Finally, we give the summary in Sec.\ref{sec:s}.

\section{Framework}
\label{sec:f}
In this section, we briefly review the two-flavor NJL model at finite isospin chemical potential. It has the Lagrangian density in terms of quark fields $\psi$~\cite{njl1,njl2,njl3,njl4,njl5,wx31}
\small{
\begin{equation}
\label{njl}
{\cal L} = \bar{\psi}\left(i \gamma^{\mu} {\partial_{\mu}}-m_{0}+\gamma^{0} \hat{\mu} \right) \psi+G\left[(\bar{\psi} \psi)^{2}+\left(\bar{\psi} i \gamma_{5} \vec{\tau} \psi\right)^{2}\right].\nonumber
\end{equation}}
Here $m_0$ is the current quark mass, $\hat{\mu}=\text{diga}(\mu_u,\mu_d)=\text{diga}(\mu_I/2,-\mu_I/2)$ is the quark chemical potential matrix with $\mu_u$ ($\mu_d$) being the $u$-quark ($d$-quark) chemical potential, and $\mu_I$ being the isospin chemical potential. The coupling constant $G$ controls the four-quark coupling strength in scalar and pseudo-scalar channels with dimension GeV$^{-2}$, and $\vec{\tau}$ is the Pauli matrix in the flavor space. At $\mu_{I}=0$, the Lagrangian density has the symmetry of $ SU_I(2) \bigotimes SU_A(2) \bigotimes U_B(1)$ corresponding to isospin symmetry, chiral symmetry, and baryon number symmetry, respectively. At $\mu_{I} \neq 0$, the $SU_I(2)$ symmetry is explicitly broken down to the global $U_I(1)$ symmetry at $\left| \mu_{I} \right| < \mu_{I}^c$, and then the $U_I(1)$ symmetry is spontaneously broken at $\left| \mu_{I} \right| > \mu_{I}^c$. Namely, the quark system enters the pion superfluid phase at the critical isospin chemical potential $\mu_{I}^c$. Note that we do not consider baryon chemical potential $\mu_B$ in this work.

\subsection{Order parameters}
Introducing two order parameters, the chiral condensate $\sigma  =\left \langle \bar \psi \psi  \right \rangle$ or effective quark mass $m=m_0-2G\sigma$ for chiral restoration phase transition and pion condensate $\Delta=-2G\left\langle\bar{\psi} i \gamma_{5} \tau_{1} \psi\right\rangle$ for pion superfluid phase transition, the inverse quark propagator matrix in the flavor space can be derived at mean field level,
\small{
\begin{equation}
\mathcal{S}_{\mathrm{mf}}^{-1}(k)=\left(\begin{array}{cc}
	\gamma^{\mu} k_{\mu}+\mu_{u} \gamma_{0}-m &  -i \Delta \gamma_{5} \\
	- i \Delta \gamma_{5} & \gamma^{\mu} k_{\mu}+\mu_{d} \gamma_{0}-m
\end{array}\right)
\label{sq}
\end{equation}}
and the thermodynamic potential of the system can be expressed as
\begin{align}\small
\Omega&( \Delta, m)=\frac{1}{4 G}\left[\left(m-m_{0}\right)^{2}+\Delta^{2}\right]-\frac{T}{V} \text{Tr} \ln \mathcal{S}_{\mathrm{mf}}^{-1},\\
&=\frac{\left(m-m_{0}\right)^{2}+\Delta^{2}}{4 G}-2 T N_{c} \int \frac{d^{3} \mathbf{k}}{(2 \pi)^{3}}\left[\ln \left(1+e^{-E_{k}^{-} / T}\right)\right. \notag \\
& \left.+\ln \left(1+e^{E_{k}^{-} / T}\right)+\ln \left(1+e^{-E_{k}^{+} / T}\right)+\ln \left(1+e^{E_{k}^{+} / T}\right)\right] . \notag
\end{align}
with quark energy $E_{k}^{ \pm} =\sqrt{\left(E_{k} \pm \mu_{I} / 2\right)^{2}+\Delta^2}$ and $E_{k}  =\sqrt{|\mathbf{k}|^{2}+m^{2}}$.

The ground state at finite temperature and isospin chemical potential is determined by minimizing the thermodynamic potential,
\begin{equation}
\frac{\partial \Omega}{\partial m}=0,\quad \frac{\partial \Omega}{\partial \Delta}=0,
\end{equation}
which give two gap equations,
\begin{align}
	\label{pioncon1}
	\Delta\left[1+4 N_{c} G \int \frac{d^{3} \mathrm{k}}{(2 \pi)^{3}}(\right. & \frac{1}{E_{k}^{-}}\left(f\left(E_{k}^{-}\right)-f\left(-E_{k}^{-}\right)\right)  \\
	& \left.+\frac{1}{E_{k}^{+}}\left(f\left(E_{k}^{+}\right)-f\left(-E_{k}^{+}\right)\right)\right]=0,\notag
\end{align}
\begin{align}
	\label{pioncon2}
	\Delta\left[1+4 N_{c} G \int \frac{d^{3} \mathrm{k}}{(2 \pi)^{3}}(\right. & \frac{1}{E_{k}^{-}}\left(f\left(E_{k}^{-}\right)-f\left(-E_{k}^{-}\right)\right)  \\
	& \left.+\frac{1}{E_{k}^{+}}\left(f\left(E_{k}^{+}\right)-f\left(-E_{k}^{+}\right)\right)\right]=0,\notag
\end{align}
with Fermi-Dirac distribution function $f(x)=(e^{x/T}+1)^{-1}$ .

Once the thermodynamic potential $\Omega$ is known, the other thermodynamic functions such as the pressure $P$, entropy density $s$, isospin density $n_I$, and energy density $\epsilon$ can be obtained through the thermodynamical relations,
\begin{equation}
	\begin{split}
		&P=-\Omega, \quad s=-\frac{\partial \Omega}{\partial T}, \quad n_{I}=-\frac{\partial \Omega}{\partial \mu_{I}},  \\
		&\epsilon=-P+T s+\mu_{I} n_{I}.
	\end{split}
\end{equation}\\

\subsection{Light mesons}

In the NJL model, the meson modes are regarded as quantum fluctuations above the quark field, and can be constructed in the frame of random phase approximation (RPA)~\cite{wx31}, where the quark bubble or polarization function is defined as

\begin{equation}
	\begin{split}
		\Pi_{M M^{\prime}}(p)=i \int \frac{d^{4} k}{(2 \pi)^{4}} \text{{Tr}} \left[\Gamma_{M}^{*} \mathcal{S}_{\mathrm{mf}}(p+k) \Gamma_{M^{\prime}} \mathcal{S}_{\mathrm{mf}}(k)\right]
	\end{split}
\end{equation}
with the vertexes
\begin{equation}
		\Gamma_{M}=\left\{\begin{array}{ll}
			1 & M=\sigma \\
			i \tau_{+} \gamma_{5} & M=\pi_{+} \\
			i \tau_{-} \gamma_{5} & M=\pi_{-} \\
			i \tau_{3} \gamma_{5} & M=\pi_{0}\  ,
		\end{array}\right.
		\Gamma_{M}^{*}=\left\{\begin{array}{ll}
			1 & M=\sigma \\
			i \tau_{-} \gamma_{5} & M=\pi_{+} \\
			i \tau_{+} \gamma_{5} & M=\pi_{-} \\
			i \tau_{3} \gamma_{5} & M=\pi_{0}\ .
		\end{array}\right.
\end{equation}

In the normal phase with explicit breaking of isospin symmetry, the eigen modes of mesons are $\sigma,\ \pi_+,\ \pi_-,\ \pi_0$. The meson mass is determined through its own pole equation
\begin{equation}
	\begin{split}
		1-2G\Pi_{MM}(p_0=M_M, {\bf p}={\bf 0})=0,
	\end{split}
\end{equation}
where we have
\begin{widetext}
	\begin{align*}
		&\Pi_{\sigma \sigma}\left(p_{0}\right)  = -  N_{c} \int \frac{d^{3} \mathbf{k}}{\left(2 \pi\right)^{3}} \frac{E_{k}^{2}-m^2}{E_{k}^2} \left(\frac{1}{E_{k}-p_{0} / 2} +\frac{1}{E_{k}+p_0/2}\right) Sign\left(E_{k}-\frac{\mu_I}{2}\right) \left(f\left(E_{k}^{+}\right)-f\left(-E_{k}^{+}\right)\right) \notag \\
		&\mathrel{\! \phantom{\Pi_{\sigma \sigma}\left(p_{0}\right)=}} - N_{c} \int \frac{d^{3} \mathbf{k}}{\left(2 \pi\right)^{3}} \frac{E_{k}^{2}-m^2}{E_{k}^2} \left(\frac{1}{E_{k}-p_{0} / 2}+\frac{1}{E_{k}+p_{0} / 2}\right) \left(f\left(E_{k}^{-}\right)-f\left(-E_{k}^{-}\right)\right),  \notag \\
		&\Pi_{\pi_{0} \pi_{0}}\left(p_{0}\right)  =- N_{c} \int \frac{d^{3} \mathbf{k}}{\left(2 \pi\right)^{3}} \left(\frac{1}{E_{k}-p_{0} / 2} +\frac{1}{E_k+p_0/2}\right) Sign\left(E_{k}-\frac{\mu_I}{2}\right)\left(f\left( E_{k}^{+}\right)-f\left(-E_{k}^{+}\right)\right) \notag \\
		&\mathrel{\! \! \phantom{\Pi_{\pi_{0} \pi_{0}}\left(p_{0}\right)=}} - N_{c} \int \frac{d^{3} \mathbf{k}}{\left(2 \pi\right)^{3}} \left(\frac{1}{E_{k}-p_{0}/2}+\frac{1}{E_k+p_0/2}\right) \left(f\left( E_{k}^{-}\right)-f\left(-E_{k}^{-}\right)\right),  \notag \\
	\end{align*}
\end{widetext}

\begin{widetext}
	\begin{align}
		\Pi_{\pi_{+} \pi_{+}}\left(p_{0}\right)  =&-2 N_{c} \int \frac{d^{3} \mathbf{k}}{\left(2 \pi\right)^{3}} \left[ \frac{1}{E_{k}+\left(p_{0}+\mu_{I}\right) / 2} \left(f\left(E_{k}^{+}\right)-f\left(-E_{k}^{+}\right)\right) +\frac{Sign \left(E_{k}-\frac{\mu_I}{2}\right)}{E_{k}-\left(p_{0}+\mu_{I}\right) / 2}\left(f\left(E_{k}^{-}\right)-f\left(-E_{k}^{-}\right)\right)\right], \notag \\
		\Pi_{\pi_{-} \pi_{-}}\left(p_{0}\right)  =&-2 N_{c} \int \frac{d^{3} \mathbf{k}}{\left(2 \pi\right)^{3}} \left[ \frac{1}{E_{k}-\left(p_{0}-\mu_{I}\right) / 2}\left(f\left(E_{k}^{+}\right)-f\left(-E_{k}^{+}\right)\right) +\frac{Sign \left(E_{k}-\frac{\mu_I}{2}\right)}{E_{k}+\left(p_{0}-\mu_{I}\right) / 2}\left(f\left(E_{k}^{-}\right)-f\left(-E_{k}^{-}\right)\right)\right],
	\end{align}
\end{widetext}
with sign function $Sign \left(E_{k}-\frac{\mu_I}{2}\right)$.

In the pion superfluid phase with spontaneous breaking of isospin symmetry, the quark propagator matrix (\ref{sq}) has non-vanishing off-diagonal terms, and leads to the mixing of meson modes. The original meson modes $\sigma,\ \pi_+,\ \pi_-,\ \pi_0$ are no longer the eigenmodes in the pion superfluid phase, and we use the notation $\bar{\sigma},\ \bar{\pi}_+, \ \bar{\pi}_-,\ \bar{\pi}_0$ for the new eigenmodes of mesons. Considering all possible channels in the bubble summation of RPA, the masses of meson modes are determined through equation
\begin{equation}
	\begin{split}
		det\left[1-2G\Pi(p_0, {\bf p}={\bf 0})\right]=0,
	\end{split}
\end{equation}
where $\Pi$ is the matrix notation for meson polarization function in $4\times 4$ meson space
	\begin{equation}
		\Pi=\left(\begin{array}{llll}
			\Pi_{\sigma \sigma} & \Pi_{\sigma \pi_{+}} & \Pi_{\sigma \pi_{-}} &  \Pi_{\sigma \pi_{0}} \\
			\Pi_{\pi_{+} \sigma} & \Pi_{\pi_{+} \pi_{+}} & \Pi_{\pi_{+} \pi_{-}} & \Pi_{\pi_{+} \pi_{0}} \\
			\Pi_{\pi_{-} \sigma} & \Pi_{\pi_{-}\pi_{+}} &  \Pi_{\pi_{-} \pi_{-}} &  \Pi_{\pi_{-} \pi_{0}} \\
			\Pi_{\pi_{0} \sigma} & \Pi_{\pi_{0} \pi_{+}} &  \Pi_{\pi_{0} \pi_{-}} &  \Pi_{\pi_{0} \pi_{0}}
		\end{array}\right) ,
	\end{equation}
and the matrix elements are written in appendix, which are almost the same as Ref~\cite{wx31} with some typos modified.

Note that the mixing between $\pi_{0}$ and other mesons disappears, because we have
\begin{equation}
	\Pi_{\pi_0 \sigma}(p_0)=\Pi_{\pi_0 \pi_+}(p_0)=\Pi_{\pi_0 \pi_-}(p_0)=0.
\end{equation}
The masses of other mesons in the pion superfluid phase are now determined by the equation
\begin{align}
	&\left[\left(\left(1-2 G \Pi_{\pi_{+} \pi_{+}}\left(p_{0}\right)\right)\left(1-2 G \Pi_{\pi_{-} \pi_{-}}\left(p_{0}\right)\right)-4 G^{2} \Pi_{\pi_{+} \pi_{-}}^{2}\left(p_{0}\right)\right)\right. \notag \\
	&\left(1-2 G \Pi_{\sigma \sigma}\left(p_{0}\right)\right)-16 G^{3} \Pi_{\sigma \pi_{+}}\left(p_{0}\right) \Pi_{\sigma \pi_{-}}\left(p_{0}\right) \Pi_{\pi_{+} \pi_{-}}\left(p_{0}\right) \notag \\
	&\mathrel{\quad \! }-4 G^{2} \Pi_{\sigma \pi_{-}}^{2}\left(p_{0}\right)\left(1-2 G \Pi_{\pi_{+} \pi_{+}}\left(p_{0}\right)\right)\notag \\
	&\mathrel{\quad \! \! }\left.-4 G^{2} \Pi_{\sigma \pi_{+}}^{2}\left(p_{0}\right)\left(1-2 G \Pi_{\pi_{-} \pi_{-}}\left(p_{0}\right)\right)\right]_{p_{0}=M_{\bar{M}}}=0 .
\end{align}
As guaranteed by the Goldstone's theorem and can be analytically proved~\cite{wx31}, there appears the massless Goldstone boson
\begin{equation}
	M_{\bar {\pi}_+}=0
\end{equation}
in the pion superfluid phase.

\section{results}
\label{sec:r}
Because of the four-fermion interaction, the NJL model is not a renormalizable theory and needs regularization. The gauge invariant Pauli-Villars regularization scheme is used in our calculations, where the quark momentum runs formally from zero to infinity~\cite{pv1,pv2,wx30,pv3}. The three parameters in the Pauli-Villars regularized NJL model, namely the current quark mass $m_0=5$ MeV, the coupling constant $G=3.44$ GeV$^{-2}$ and the Pauli-Villars mass parameter $\Lambda=1127$ MeV are fixed by fitting the chiral condensate $\langle\bar\psi\psi\rangle=-(250\ \text{MeV})^3$, pion mass $m_\pi=134$ MeV and pion decay constant $f_\pi=93$ MeV in vacum with $T=\mu_B=\mu_I=0$. The color degrees of freedom in the NJL model is trivial, and the trace in color space simply contributes a factor $N_{c}=3$.

In order to study the coupling strength effect, we vary the parameter $G$ in our calculations and keep other parameters intact. We find that the increase of coupling strength triggers a BCS-BEC crossover along the phase boundary of pion superfluid phase transition. In the following calculations, we consider the situation with $T=\mu_B=0$ and $\mu_I>0$.

\subsection{Phase structure}
\label{seca}
At vanishing temperature, the two gap equations (\ref{pioncon1}) and (\ref{pioncon2}) become
\small{
\begin{eqnarray}
 \int \frac{d^{3} \mathbf{k}}{(2 \pi)^{3}} \frac{m }{E_{k}}\left(\frac{E_{k}-\mu_{I} / 2}{E_{k}^{-}}+\frac{E_{k}+\mu_{I} / 2}{E_{k}^{+}}\right)&=&\frac{m-m_0}{4GN_c},
\label{mq}\\
\Delta\left[1-4N_c G \int \frac{d^{3} \mathbf{k}}{(2 \pi)^{3}}\left(\frac{1}{E_{k}^{-}}+\frac{1}{E_{k}^{+}}\right)\right]&=&0.
\label{con}
\end{eqnarray}
}

\begin{figure*}[hbt]
	\centering
	\begin{minipage}{0.49\textwidth}
		\centering
		\includegraphics[width=0.9\textwidth]{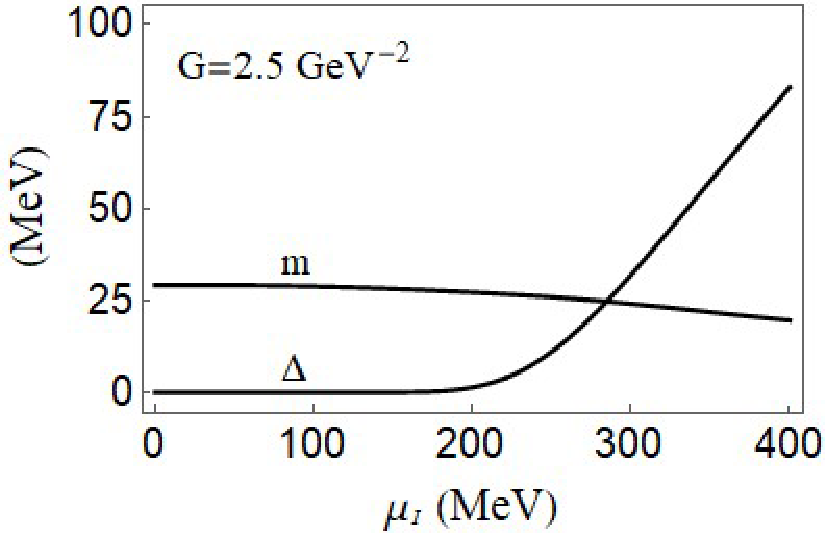}
	\end{minipage}
	\begin{minipage}{0.49\textwidth}
		\centering
		\includegraphics[width=0.9\textwidth]{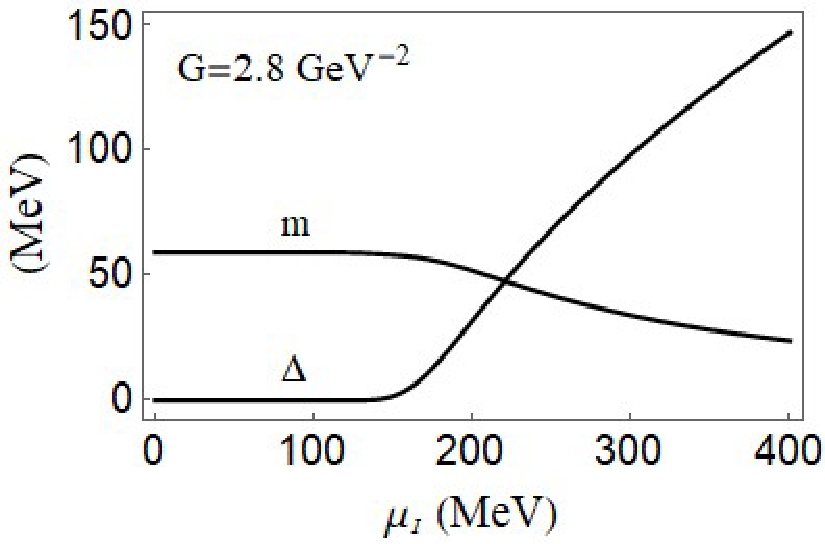}
	\end{minipage}
	\qquad
	\begin{minipage}{0.49\textwidth}
		\centering
		\includegraphics[width=0.9\textwidth]{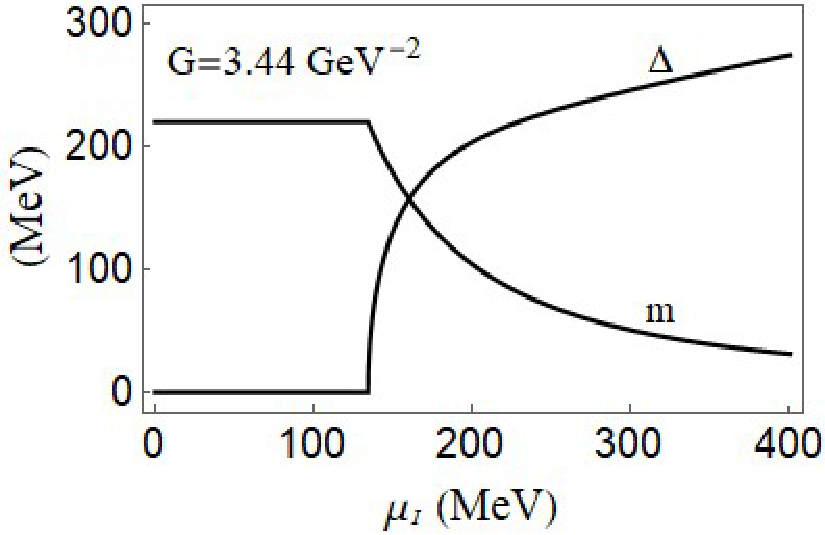}
		\end{minipage}
	\begin{minipage}{0.49\textwidth}
		\centering
		\includegraphics[width=0.9\textwidth]{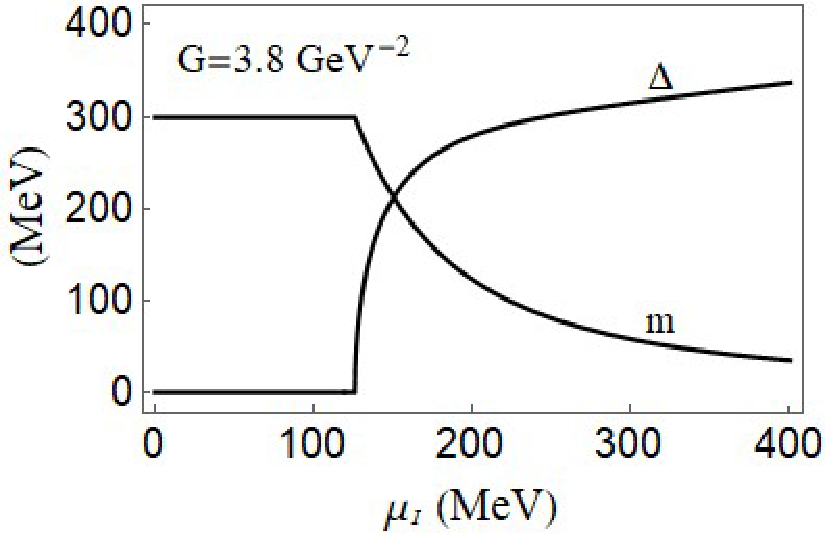}
	\end{minipage}
	\caption{The pion condensate $\Delta$ and effective quark mass $m$ as functions of isospin chemical potential $\mu_I$ with different coupling strength $G=2.5,\ 2.8,\ 3.44,\ 3.8$ GeV$^{-2}$.}
\label{cons}
\end{figure*}

At low isospin chemical potential region, the order parameter pion condensate $\Delta=0$ is always a solution of Eq.(\ref{con}), which corresponds to the normal phase without spontaneous breaking of isospin symmetry. Meanwhile, the other order parameter effective quark mass $m$ is isospin chemical potential $\mu_I$ independent,
\begin{equation}
\frac{m_0-m}{2G} +4 N_{c} \int \frac{d^{3} \mathbf{k}}{(2 \pi)^{3}} \frac{m }{E_{k}}=0,
\label{plat}
\end{equation}
when $m>\frac{\mu_I}{2}$.

At the critical isospin chemical potential $\mu_I^c$, where the isospin symmetry starts to break spontaneously and the pion condensate begins to appear, the solution $\Delta=0$ should satisfies the equation
\begin{equation}
1-4 N_{c} G \int \frac{d^{3} \mathbf{k}}{(2 \pi)^{3}}\left(\frac{1}{E_{k}^{-}}+\frac{1}{E_{k}^{+}}\right)=0.
\end{equation}
When $m>\frac{\mu_I}{2}$, it can be rewritten as
\begin{equation}
1-8 N_{c} G \int \frac{d^{3} \mathbf{k}}{(2 \pi)^{3}}\frac{E_{k}}{E_{k}^2-\left(\mu^c_I\right)^2/4}=0.
\end{equation}
From the comparision with the well-known pole equation~\cite{njl3,wx36,wx37} determining the pion mass in vacuum $M_\pi$,
\begin{equation}
1-8 N_{c} G \int \frac{d^{3} \mathbf{k}}{(2 \pi)^{3}}\frac{E_{k}}{E_{k}^2-\left(M_\pi\right)^2/4}=0,
\label{piv}
\end{equation}
we find that the critical isospin chemical potential $\mu^c_I$ for the pion superfluid phase transition is exactly equal to the pion mass in vacuum
\begin{equation}
\mu^c_I=M_\pi.
\label{muimpi}
\end{equation}
However, when $m<\frac{\mu_I}{2}$, no such conclusions can be obtained. It should be mentioned that the above derivations do not depend on the regularization schemes.

Fig.\ref{cons} shows the pion condensate $\Delta$ and the effective quark mass $m$ as functions of isospin chemical potential $\mu_I$ with different coupling strength $G=2.5,\ 2.8,\ 3.44,\ 3.8$ GeV$^{-2}$. With fixed coupling strength, the pion condensate continuously increases from zero to nonzero value with increasing isospin chemical potential. The pion superfluid phase transition at the critical isospin chemical potential $\mu_{I}^c$ is of second order, and we have $\mu_{I}^c=128,\ 123,\ 134,\ 126$ MeV at $G=2.5,\ 2.8,\ 3.44,\ 3.8$ GeV$^{-2}$, respectively. For weak (strong) coupling cases $G=2.5,\ 2.8$ GeV$^{-2}$ ($G=3.44,\ 3.8$ GeV$^{-2}$), we observe a slow (fast) increasing slope of pion condensate around $\mu_{I}^c$ and a small (large) value of pion condensate at high isospin chemical potential. The effective quark mass $m$ is also sensitive to the coupling strength $G$, which shows a larger value with increasing $G$. In strong coupling cases $G=3.44,\ 3.8$ GeV$^{-2}$, a platform structure of effective quark mass appears in the normal phase at the low $\mu_I$ region, as expected from Eq.(\ref{plat}). But in weak coupling cases $G=2.5,\ 2.8$ GeV$^{-2}$, the platform structure disappears. The effective quark mass has a slow (fast) decreasing slope in the weak (strong) coupling pion superfluid phase. 

\begin{figure}[hbt]
	\centering
	\includegraphics[width=8cm]{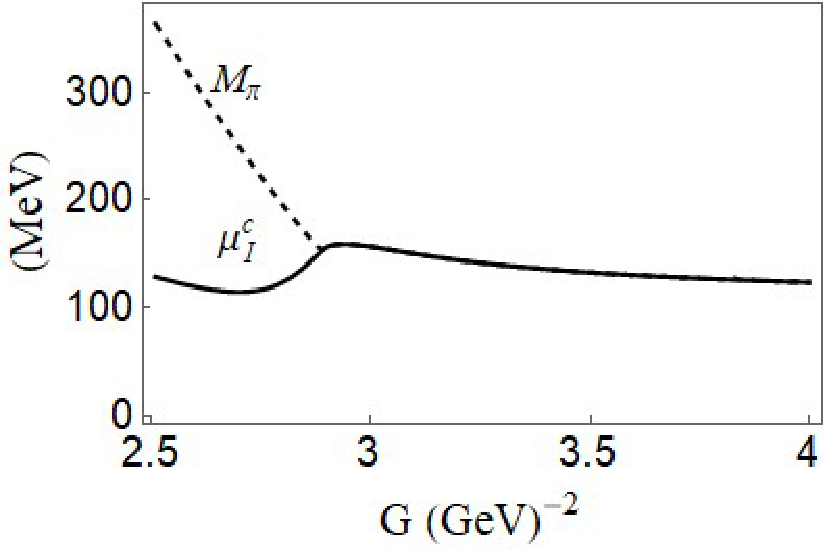}
	\caption{The critical isospin chemical potential $\mu_{I}^c$ and pion mass in vacuum $M_\pi$ as functions of coupling strength $G$. They become exactly same when $G\geq G_0 = 2.89$ GeV$^{-2}$.}
	\label{mui1}
\end{figure}

Fig.\ref{mui1} shows the critical isospin chemical potential $\mu_{I}^c$ for pion superfluid phase transition, and pion mass in vacuum $M_\pi$, as functions of coupling strength $G$. The critical isospin chemical potential $\mu_{I}^c$ is a non-monotonic function of coupling strength $G$. With increasing $G$, $\mu_{I}^c$ first decreases, then increases and finally decreases. With $G\geq G_0=2.89$ GeV$^{-2}$, we solve a large value of quark mass $m>\mu_I/2$ in the normal phase, and therefore the critical isospin chemical potential $\mu_{I}^c$ is exactly the same as pion mass $M_\pi$ in vacuum, as derived in Eq.(\ref{muimpi}). With $G<G_0$, which gives a small value of quark mass $m<\mu_I/2$ in the normal phase, we have $\mu_{I}^c<M_\pi$.

\begin{figure}[hbt]
	\centering
	\includegraphics[width=8cm]{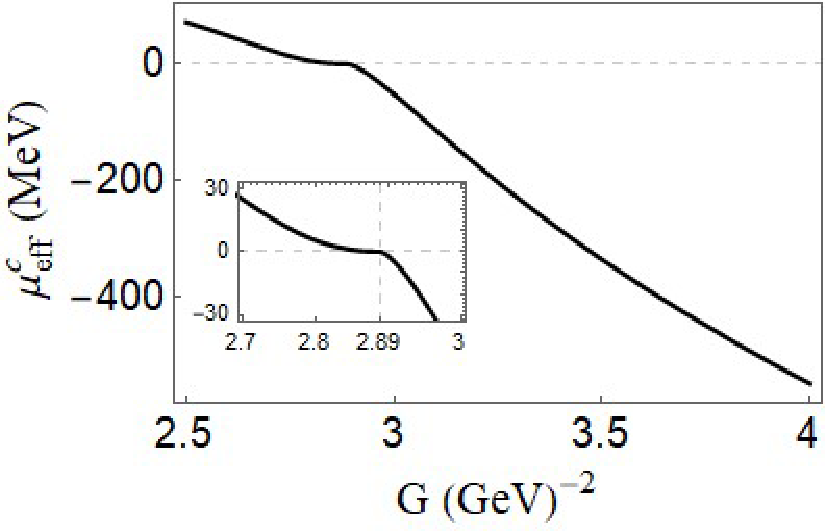}
	\caption{The effective chemical potential $\mu_{\text{eff}}^c$ at the critical point of pion superfluid phase transition as a function of coupling strength $G$. $\mu_{\text{eff}}^c=0$ is obtained at $G=G_0 = 2.89$ GeV$^{-2}$.}
	\label{mui2}
\end{figure}

In relativistic systems, the effective chemical potential is defined as
\begin{equation}
\frac{\mu_{\text{eff}}}{2} =\frac{ \mu_I}{2} - m,
\end{equation}
and $\mu_{\text{eff}} =0$ is a measure of the BCS-BEC crossover in the superfluid phase, with $\mu_{\text{eff}}>0$ in BCS state and $\mu_{\text{eff}}<0$ in BEC state. In Fig.\ref{mui2}, we plot the effective chemical potential at the critical point of pion superfluid phase transition $\mu_{\text{eff}}^c=\mu_{\text{eff}}(\mu_{I}^c)$ as a function of the coupling strength $G$. We observe that $\mu_{\text{eff}}^c=0$ happens at $G=G_0$, and $\mu_{\text{eff}}^c>0$ ($\mu_{\text{eff}}^c<0$) at $G<G_0$ ($G>G_0$). A BCS-BEC crossover along the phase boundary of pion superfluid phase transition is induced by the increasing coupling strength between quarks. The pion superfluid phase around $\mu_{I}^c$ is in the BEC state with $G\geq G_0$, which is accompanied with a fast increase of pion condensate, and in BCS state with $G<G_0$, which is accompanied with a slow increase of pion condensate.

To summary, in coupling strength and isospin chemical potential plane, the pion superfluid phase structure can be understood as follows. In strong coupling cases $G\geq G_0$, the quark matter first undergoes BEC state after entering the pion superfluid phase and then becomes BCS state at higher isospin chemical potential. In this case, the critical isospin chemical potential for the pion superfluid phase transition is exactly equal to the pion mass in vacuum $\mu^c_I=M_\pi$. In weak coupling cases $G<G_0$, the quark matter only undergoes BCS state in the pion superfluid phase, and we obtain $\mu^c_I<M_\pi$.


\subsection{Meson mass}
\label{secmeson}
\begin{figure}[hbt]
	\centering
	\includegraphics[width=8cm]{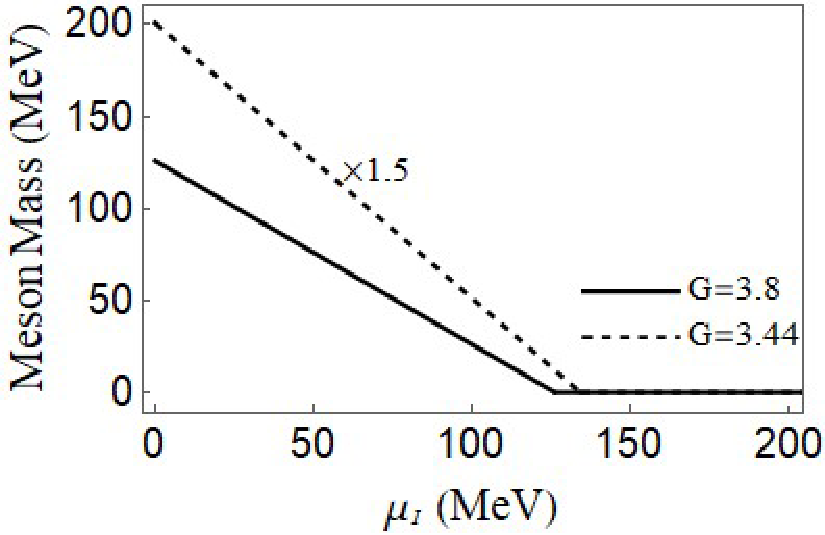}
	\includegraphics[width=8cm]{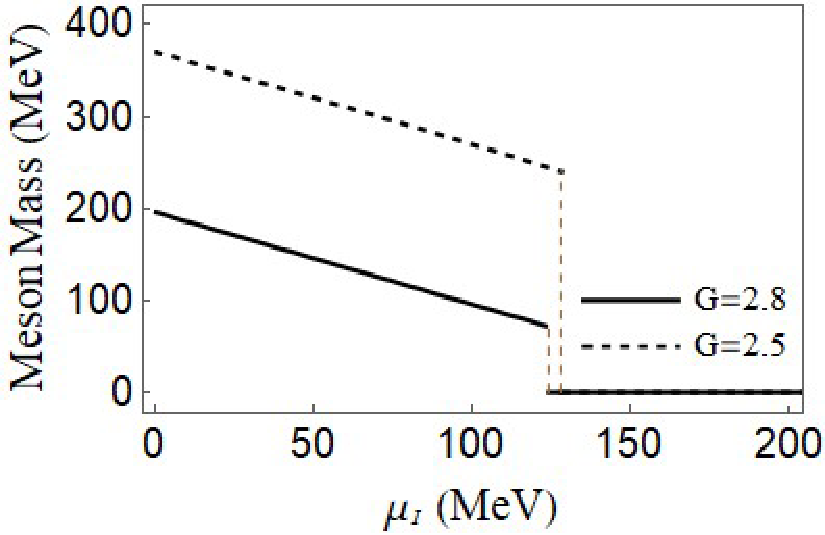}
	\caption{The mass of $\pi_{+}$ ($\bar{\pi}_{+}$) meson as a function of isospin chemical potential $\mu_I$ with different coupling strength $G= 3.44,\ 3.8$ GeV$^{-2}$ (upper panel) and $G=2.5,\ 2.8$ GeV$^{-2}$ (lower panel). In the upper panel, the meson masses with $G=3.44$ GeV$^{-2}$ and $G=3.8$ GeV$^{-2}$ has similar values. For ease of reading, the mass with $G=3.44$ GeV$^{-2}$ is plotted as 1.5 times of the original value.}
	\label{mesonmassdata}
\end{figure}

The mechanism for a continuous phase transition is the spontaneous symmetry breaking. One can define the order parameter which changes from nonzero value to zero or vice verse when the phase transition happens, see our discussion in the previous section. On the other hand, the spontaneous breaking of a global symmetry manifests itself in the emergence of massless Goldstone bosons, based on Goldstone's theorem~\cite{wx38,wx39}. For the pion superfluid phase transition, the Goldstone mode is the $\bar{\pi}_{+}$ meson. In this part, we discuss the coupling strength effect on the mass of $\pi_{+}$ ($\bar{\pi}_{+}$) meson, and Fig.\ref{mesonmassdata} depicts the mass of $\pi_{+}$ ($\bar{\pi}_{+}$) meson as a function of isospin chemical potential $\mu_I$ with different coupling strength $G=2.5,\ 2.8,\ 3.44,\ 3.8$ GeV$^{-2}$.

At vanishing temperature, the $\pi_{+}$ meson polarization function in normal phase becomes
\small{
\begin{eqnarray}
	&&\Pi_{\pi_{+} \pi_{+}}\left(p_{0}\right)  =\\ \nonumber &&2 N_{c} \int \frac{d^{3} \mathbf{k}}{\left(2 \pi\right)^{3}} \left( \frac{1}{E_{k}+\frac{p_{0}+\mu_{I}}{2}} +\frac{{Sign} \left(E_{k}-\frac{\mu_I}{2}\right)}{E_{k}-\frac{p_{0}+\mu_{I}}{2}}\right).
\end{eqnarray}}

When $m>\frac{\mu_I}{2}$, the pole equation of $\pi_{+}$ meson can be simplified as
\begin{equation}
	1-8 N_{c} G \int \frac{d^{3} \mathbf{k}}{\left(2 \pi\right)^{3}}  \frac{E_{k}}{E_{k}^2-\left(p_{0}+\mu_{I}\right)^2 /4}=0.
\end{equation}
Making comparison with the pole equation (\ref{piv}) for pion mass in vacuum $M_\pi$, we have
\begin{equation}
M_{\pi_+}=M_\pi-\mu_I,
\end{equation}
in the normal phase. Therefore, at the critical point $\mu_I=\mu_I^c$, we have
\begin{equation}
M_{\pi_+}=0,
\end{equation}
which is smoothly connected with the massless Goldstone mode $\bar{\pi}_{+}$ in the pion superfluid phase. Such results are obtained in strong coupling cases $G \geq G_0$, see Fig.\ref{mesonmassdata} with $G=3.44,\ 3.8$ GeV$^{-2}$ as examples.

With weak coupling strength $G<G_0$, we have $m<\frac{\mu_I}{2}$ in the normal phase, no such analytic derivations can be obtained. We have to do numerical calculations to solve $M_{\pi_+}$, see examples in Fig.\ref{mesonmassdata} with $G=2.5,\ 2.8$ GeV$^{-2}$. $M_{\pi_+}$ is a decreasing function of isospin chemical potential. Approximately, it also obeys the relation $M_{\pi_+}=M_\pi-\mu_I$. Since in this case, the critical isospin chemical potential $\mu_I^c$ is smaller than pion mass in vacuum $M_\pi$, a mass jump of $M_{\pi_+}$ appears at the critical point $\mu_I=\mu_I^c$.

It should be mentioned that when solving the mass spectra of mesons, we find that the meson mixing is also sensitive to the coupling strength $G$, which is not explained in detail here.

\subsection{Mass-radius relation}
\label{secb}
Compact stars are the possible laboratory for quark matter with finite isospin density, and the structure of compact stars is closely related to the equation of state $P(\epsilon)$ of quark matter inside.

For a non-rotating and spherically symmetric star, its mass and radius are determined by the Tolman-Oppenheimer-Volkoff (TOV) equations~\cite{wx40,wx41}
\begin{equation}
	\begin{split}
		\frac{d P}{d r} & =-\frac{G_{N}(\epsilon+P)\left(M+4 \pi r^{3} P\right)}{r^{2}\left(1-2 G_{N} M / r\right)}, \\
		\frac{d M}{d r} & =4 \pi r^{2} \epsilon,
	\end{split}
\end{equation}
where $P(r)$ and $\epsilon(r)$ are the pressure and energy density at radius $r$ inside the star, respectively, and $M(r)$ is the total mass contained within a sphere of radius $r$. Substituting an equation of state $P(\epsilon)$ and giving a fixed central pressure $P_c$, the star mass and radius can be numerically solved by integrating the TOV equations from the center of the star up to its surface $r=R$, where the pressure reaches its perturbative value $P(R)=B$ with the MIT bag constant $B=75$ MeV$\cdot$fm$^{-3}$~\cite{pv3,wx42}.

\begin{figure}[hbt]
	\centering
	\includegraphics[width=8cm]{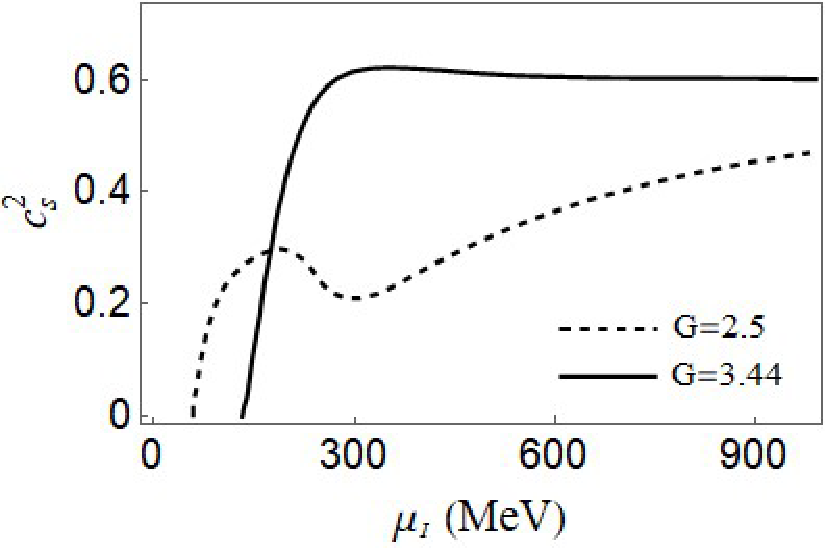}
	\caption{The sound velocity $c_s^2=\frac{\partial P}{\partial \epsilon}$ as a function of isospin chemical potential with different coupling strength $G=2.5,\ 3.44$ GeV$^{-2}$.}
	\label{eos}
\end{figure}

Figure \ref{eos} plots the sound velocity $c^2_s=\frac{\partial P}{\partial \epsilon}$ of quark matter as a function of isospin chemical potential with different coupling strength $G=2.5,\ 3.44$ GeV$^{-2}$. In case of strong coupling $G=3.44$ GeV$^{-2}$, the sound velocity $c^2_s$ increases fast around the critical isospin chemical potential due to the appearance of BEC state in pion superfluid phase, and then becomes saturate $c^2_s\simeq 0.6$ at high isospin chemical potential. For weak coupling case $G=2.5$ GeV$^{-2}$, the quark matter is in BCS state of pion superfluid phase. $c^2_s$ increases slowly and non-monotonically, and we obtain $c^2_s\simeq 0.4$ at high isospin chemical potential. With stronger coupling between quarks, we have a stiffer EoS (larger $c^2_s$) for quark matter, which will lead to a heavier mass and larger radius for the corresponding compact star.

\begin{figure}[hbt]
	\centering
	\includegraphics[width=8cm]{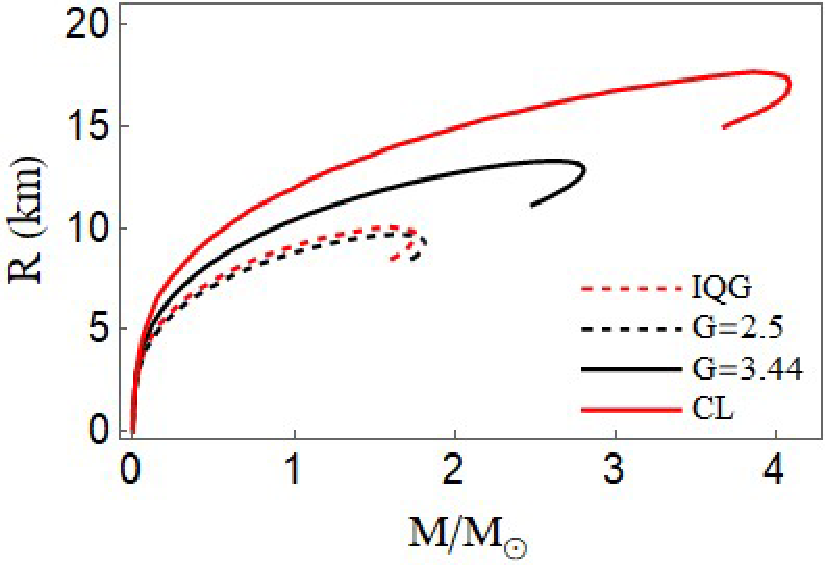}
	\caption{The mass-radius relation of compact stars composed of pion superfluid quark matter with different coupling strength $G=2.5,\ 3.44$ GeV$^{-2}$. CL and IQG mean quark matter in the causal limit with sound velocity $c_s^2=1$ and idea quark gas with $c_s^2=1/3$, respectively.}
	\label{mr}
\end{figure}

Figure \ref{mr} shows the mass-radius relation for compact stars composed of pion superfluid quark matter with different coupling strength $G=2.5,\ 3.44$ GeV$^{-2}$. For comparison, we also calculate the mass-radius relation in cases of idea quark gas (IQG) with $c^2_s=1/3$ and causal limit (CL) with $c^2_s=1$. In case of weak coupling pion superfluid quark matter, the resulting mass-radius relation is similar with the idea quark gas, due to its soft EoS. The maximum star mass and radius can reach $M\sim 2 M_\odot$ and $R \sim 9$ km, respectively. For the strong coupling case, the stiff EoS leads to the larger values of maximum star mass $M\sim3 M_\odot$ and radius $R \sim 14$ km. Referring to the measured data, the weak coupling pion superfluid quark matter may exist in the observed compact stars, such as PSR J1614-2230, PSR J0030+0451, PSR J1903+0327 and PSR J0740+6620~\cite{wx43,wx44,wx45,wx46}, and the strong coupling pion superfluid quark matter maybe a candidate to explain the massive compact stars, such as PSR J1311-3430, PSR J0952-0607, GW170817and GW190814~\cite{wx47,romani2022psr,abbott2017gw170817,dexheimer2021gw190814}.

\section{summary}
\label{sec:s}
Coupling strength effect on the quark matter with finite isospin chemical potential is investigated by a Pauli-Villars regularized NJL model. When fixing coupling strength $G$, the pion superfluid phase transition happens at a critical isospin chemical potential $\mu_I^c$, which is a non-monotonic function of $G$. A BCS-BEC crossover along the phase boundary of pion superfluid phase transition is triggered by the increasing coupling strength. For strong coupling cases $G\geq G_0$, the $\mu_I^c$ is exactly the same as pion mass in vacuum $M_\pi$. At the critical point $\mu_I=\mu_I^c$, both the order parameters and mass of Goldstone mode change continuously. Around $\mu_I^c$, the pion superfluid quark matter is in BEC state, associated with a fast increase of pion condensate. For weak coupling cases $G< G_0$, we obtain $\mu_I^c<M_\pi$, and a mass jump of Goldstone boson. The pion superfluid quark matter is in BCS state even around the critical point $\mu_I^c$, accompanied by a slow increase of pion condensate.

Coupling strength effect also causes the change of bulk properties of quark matter. In strong (weak) coupling cases, the EoS of quark matter at finite isospin chemical potential is stiff (soft). The compact stars composed of strong (weak) coupling pion superfluid quark matter has a heavier (lighter) mass and larger (smaller) radius. Referring to the measured data, the weak coupling pion superfluid quark matter may exist in the observed compact stars, such as PSR J1614-2230, PSR J0030+0451, PSR J1903+0327 and PSR J0740+6620, and the strong coupling pion superfluid quark matter maybe a candidate to explain the massive compact stars like PSR J1311-3430, PSR J0952-0607, GW170817and GW190814.

In this work, we do not consider the baryon chemical potential $\mu_B$, which serves as the Fermi surface mismatch between $u$-quark and $\bar d$-quark. As shown in the previous study~\cite{wx48,wx49,wx50,wx51,pv3,mucf}, $\mu_B$ may induce some exotic superfluid states, such as gapless Sarma state and inhomogeneous LOFF state, near the phase boundary in $\mu_I-\mu_B$ plane. The coupling strength effect on these exotic superfluid states is under progress and will be reported in the near future.

\section{Appendix}

The meson polarization functions in the basis ($\sigma$,$\pi_+$,$\pi_-$,$\pi_0$) at $\bf p=0$ are defined as

\begin{widetext}

	\begin{align*}
		&\Pi_{\pi_0 \pi_0}\left(p_{0}\right)= 2 N_{c} \int \frac{d^{3} \mathbf{k}}{(2 \pi)^{3}}\left[\frac { E _ { k } ^ { - } E _ { k } ^ { + } - ( E _ { k } ^ { 2 } - \mu _ { I } ^ { 2 } / 4 ) - \Delta^2 } { p _ { 0 } ^ { 2 } - ( E _ { k } ^ { - } - E _ { k } ^ { + } ) ^ { 2 } } ( \frac { 1 } { E _ { k } ^ { + } } - \frac { 1 } { E _ { k } ^ { - } } ) \left(f\left(E_{k}^{-}\right)+f\left(-E_{k}^{+}\right)-f\left(-E_{k}^{-}\right)-f\left(E_{k}^{+}\right)\right)\right. \notag \\
		&\mathrel{\! \! \phantom{\Pi_{\pi_0 \pi_0}\left(p_{0}\right)}} \left.+\frac{E_{k}^{-} E_{k}^{+}+\left(E_{k}^{2}-\mu_{I}^{2} / 4\right)+\Delta^2}{p_{0}^{2}-\left(E_{k}^{-}+E_{k}^{+}\right)^{2}}\left(\frac{1}{E_{k}^{+}}+\frac{1}{E_{k}^{-}}\right)\left(f\left(E_{k}^{-}\right)+f\left(E_{k}^{+}\right)-f\left(-E_{k}^{-}\right)-f\left(-E_{k}^{+}\right)\right)\right], \\
		&\Pi_{\sigma \sigma}\left(p_{0}\right)= 2 N_{c} \int \frac{d^{3} \mathbf{k}}{(2 \pi)^{3}} \frac{E_k^2-m^2}{E_{k}^{2}}\left[\frac{E_{k}^{-} E_{k}^{+}-\left(E_{k}^{2}-\mu_{I}^{2} / 4\right)-\Delta^2}{p_{0}^{2}-\left(E_{k}^{-}-E_{k}^{+}\right)^{2}}\left(\frac{1}{E_{k}^{+}}-\frac{1}{E_{k}^{-}}\right)\left(f\left(E_{k}^{-}\right)+f\left(-E_{k}^{+}\right)-f\left(-E_{k}^{-}\right)-f\left(E_{k}^{+}\right)\right)\right. \notag \\
		&\mathrel{\! \! \phantom{\Pi_{\sigma \sigma}\left(p_{0}\right)}} \left.\left.+\frac{E_{k}^{-} E_{k}^{+}+\left(E_{k}^{2}-\mu_{I}^{2} / 4\right)+\Delta^2}{p_{0}^{2}-\left(E_{k}^{-}+E_{k}^{+}\right)^{2}}\left(\frac{1}{E_{k}^{+}}+\frac{1}{E_{k}^{-}}\right)\left(f\left(E_{k}^{-}\right)+f\left(E_{k}^{+}\right)-f\left(-E_{k}^{-}\right)-f(-E_{k}^{+}\right)\right)\right] \notag \\
		&\mathrel{ \phantom{\Pi_{\sigma \sigma}\left(p_{0}\right)}} +8 N_{c} \int \frac{d^{3} \mathbf{k}}{(2 \pi)^{3}} \frac{m^{2}}{E_{k}^{2}}\left[\frac{\Delta^2}{p_{0}^{2}-4\left(E_{k}^{-}\right)^{2}} \frac{1}{E_{k}^{-}}\left(f\left(E_{k}^{-}\right)-f\left(-E_{k}^{-}\right)\right)+\frac{\Delta^2}{p_{0}^{2}-4\left(E_{k}^{+}\right)^{2}} \frac{1}{E_{k}^{+}}\left(f\left(E_{k}^{+}\right)-f\left(-E_{k}^{+}\right)\right)\right], \\
		&\Pi_{\pi_{+} \pi_{-}}\left(p_{0}\right)= \Pi_{\pi_{-} \pi_{+}}\left(p_{0}\right) \notag \\
		&\mathrel{ \phantom{\Pi_{\pi_{+} \pi_{-}}\left(p_{0}\right)}}=  -4 N_{c} \int \frac{d^{3} \mathbf{k}}{(2 \pi)^{3}}\left[\frac{\Delta^2}{p_{0}^{2}-4\left(E_{k}^{-}\right)^{2}} \frac{1}{E_{k}^{-}}\left(f\left(E_{k}^{-}\right)-f\left(-E_{k}^{-}\right)\right)+\frac{\Delta^2}{p_{0}^{2}-4\left(E_{k}^{+}\right)^{2}} \frac{1}{E_{k}^{+}}\left(f\left(E_{k}^{+}\right)-f\left(-E_{k}^{+}\right)\right)\right], \\
		&\Pi_{\sigma \pi_{+}}\left(p_{0}\right)=  \Pi_{\pi_{+} \sigma}\left(p_{0}\right) \notag \\
		&\mathrel{ \phantom{\Pi_{\sigma \pi_{+}}\left(p_{0}\right)}}= -2 \sqrt{2} N_{c} \Delta \int \frac{d^{3} \mathbf{k}}{(2 \pi)^{3}} \frac{m}{E_{k}}\left[\frac{2 E_{k}-\mu_{I}+p_{0}}{p_{0}^{2}-4\left(E_{k}^{-}\right)^{2}} \frac{1}{E_{k}^{-}}\left(f\left(E_{k}^{-}\right)-f\left(-E_{k}^{-}\right)\right)\right. \left.+\frac{2 E_{k}+\mu_{I}-p_{0}}{p_{0}^{2}-4\left(E_{k}^{+}\right)^{2}} \frac{1}{E_{k}^{+}}\left(f\left(E_{k}^{+}\right)-f\left(-E_{k}^{+}\right)\right)\right], \\
		&\Pi_{\sigma \pi_{-}}\left(p_{0}\right)= \Pi_{\pi_{-} \sigma}\left(p_{0}\right) \notag \\
		&\mathrel{ \phantom{\Pi_{\sigma \pi_{-}}\left(p_{0}\right)}}= -2 \sqrt{2} N_{c} \Delta \int \frac{d^{3} \mathbf{k}}{(2 \pi)^{3}} \frac{m}{E_{k}}\left[\frac{2 E_{k}-\mu_{I}-p_{0}}{p_{0}^{2}-4\left(E_{k}^{-}\right)^{2}} \frac{1}{E_{k}^{-}}\left(f\left(E_{k}^{-}\right)-f\left(-E_{k}^{-}\right)\right)\right. \left.+\frac{2 E_{k}+\mu_{I}+p_{0}}{p_{0}^{2}-4\left(E_{k}^{+}\right)^{2}} \frac{1}{E_{k}^{+}}\left(f\left(E_{k}^{+}\right)-f\left(-E_{k}^{+}\right)\right)\right],\\
		&\Pi_{\pi_{-} \pi_{-}}\left(p_{0}\right)= 4 N_{c} \int \frac{d^{3} \mathbf{k}}{(2 \pi)^{3}}\left[\frac{\left(E_{k}^{-}\right)^{2}+\left(E_{k}-\mu_{I} / 2\right)^{2}-p_{0}\left(E_{k}-\mu_{I} / 2\right)}{p_{0}^{2}-4\left(E_{k}^{-}\right)^{2}} \frac{1}{E_{k}^{-}}\left(f\left(E_{k}^{-}\right)-f\left(-E_{k}^{-}\right)\right)\right. \notag \\
		&\mathrel{ \qquad \qquad \qquad \phantom{\Pi_{\pi_{-} \pi_{-}}\left(p_{0}\right)=}} \left.+\frac{\left(E_{k}^{+}\right)^{2}+\left(E_{k}+\mu_{I} / 2\right)^{2}+p_{0}\left(E_{k}+\mu_{I} / 2\right)}{p_{0}^{2}-4\left(E_{k}^{+}\right)^{2}} \frac{1}{E_{k}^{+}}\left(f\left(E_{k}^{-}\right)-f\left(-E_{k}^{+}\right)\right)\right],\\
		&\Pi_{\pi_{+} \pi_{+}}\left(p_{0}\right)= 4 N_{c} \int \frac{d^{3} \mathbf{k}}{(2 \pi)^{3}}\left[\frac{\left(E_{k}^{-}\right)^{2}+\left(E_{k}-\mu_{I} / 2\right)^{2}+p_{0}\left(E_{k}-\mu_{I} / 2\right)}{p_{0}^{2}-4\left(E_{k}^{-}\right)^{2}} \frac{1}{E_{k}^{-}}\left(f\left(E_{k}^{-}\right)-f\left(-E_{k}^{-}\right)\right)\right. \notag \\
		&\mathrel{ \qquad \qquad \qquad \phantom{\Pi_{\pi_{+} \pi_{+}}\left(p_{0}\right)=}} \left.+\frac{\left(E_{k}^{+}\right)^{2}+\left(E_{k}+\mu_{I} / 2\right)^{2}-p_{0}\left(E_{k}+\mu_{I} / 2\right)}{p_{0}^{2}-4\left(E_{k}^{+}\right)^{2}} \frac{1}{E_{k}^{+}}\left(f\left(E_{k}^{+}\right)-f\left(-E_{k}^{+}\right)\right)\right].
	\end{align*}
\end{widetext}

\noindent {\bf Acknowledgement:}
The work is supported by the NSFC Grant 11775165.

\end{document}